\documentclass[runningheads]{llncs}
\usepackage{graphicx}
\usepackage{multirow}
\usepackage{colortbl}
\usepackage{caption}
\usepackage{subcaption}
\usepackage{hyperref}
\usepackage{amssymb}
\usepackage{amsmath}
\begin{document}
\title{You can't see what you can't see: Experimental evidence for how much relevant information may be missed due to Google's Web search personalisation}
\titlerunning{You can't see what you can't see}
%
\author{Cameron Lai\inst{1} \and
Markus Luczak-Roesch\inst{1}\orcidID{0000-0003-4610-7244}}
\authorrunning{C. Lai and M. Luczak-Roesch}
%
\institute{Victoria University of Wellington \\
School of Information Management \\
Wellington, New Zealand \\
\email{Markus.Luczak-Roesch@vuw.ac.nz}\\
\url{https://www.victoria.ac.nz/sim}}
\maketitle              
\begin{abstract}
The influence of Web search personalisation on professional knowledge work is an understudied area. Here we investigate how public sector officials self-assess their dependency on the Google Web search engine, whether they are aware of the potential impact of algorithmic biases on their ability to retrieve all relevant information, and how much relevant information may actually be missed due to Web search personalisation. We find that the majority of participants in our experimental study are neither aware that there is a potential problem nor do they have a strategy to mitigate the risk of missing relevant information when performing online searches. Most significantly, we provide empirical evidence that up to $20\%$ of relevant information may be missed due to Web search personalisation. This work has significant implications for Web research by public sector professionals, who should be provided with training about the potential algorithmic biases that may affect their judgments and decision making, as well as clear guidelines how to minimise the risk of missing relevant information.

\end{abstract}
\section{Introduction}
In August 2018 Radio New Zealand (RNZ), a New Zealand public radio broadcaster, reported that the use of Google Web search by the New Zealand Police may have unwittingly revealed a link between two suspects facing charges for a crime committed together but who had no documented history of any joint crime or crime of the same kind \cite{rnz2018}. This particular incident was significant because one of the suspects had official name suppression, so that the revelation of the name could have opened a loophole for the defending lawyers to counter the charges on the basis of the violation of the suspect's rights. So far it is assumed that the police, when investigating the two suspects by performing Web searches using Google, triggered the search engine's algorithms to learn a connection, which led to them appearing together in the Google search results when searching for the one suspect only whose name was not officially suppressed. 

Because the inner workings of the Google search ranking and personalisation algorithms are likely to remain a corporate secret \cite{ormen2016googling}, it can only be speculated about what caused this particular incident to happen. However, the case highlights a general and important issue with public sector officials' work that involves the use of digital services provided by global IT companies. Some of the questions in this problem domain, which we suggest is as an understudied area that requires timely and deeper investigation, are: What is the impact of search engine personalisation on the work of public sector officials? Which technical features of search engine personalisations impact public sector officials' work? Is it possible for public sector officials to prevent being affected by search engine personalisation with respect to their work? 

Here we contribute to this line of scientific inquiry by performing an experiment involving public sector professionals from a range of governmental agencies in New Zealand. In order to understand the impact of Google's search result personalisation on knowledge work in the public sector, we address the following questions: (RQ1) How reliant are public sector officials on the use of Google search? (RQ2) Is there a difference between personalised and un-personalised Google search for queries in different public sector agencies? (RQ3) How does the personalisation of search results affect the perceived relevance of search results for public sector officials with respect to their work?

By answering these questions we make the following contributions: First, we show how highly public sector officials self-assess their dependency on Google Web search and provide evidence for a lack of awareness that Google search personalisation may have an impact on knowledge work in professional contexts. Second, we quantify the amount of relevant information that may be missed due to Web search personalisation. Third, we provide insight into how alternative search practices may help to overcome this issue.

The remainder of this paper is structured as follows: We begin with a description of the foundations of Web search and Web search personalisation, followed by a review of related studies that looked into quantifying the impact of Web search personalisation. Informed by the related studies, we describe the research design and subsequent results. We then discuss the implications of the results for research and practice.

\section{Preliminaries and related work}

\subsection{A brief history of Web search}

Yahoo, AltaVista, Lycos and Fireball were among the first search engines to emerge when the World Wide Web was established \cite{holscher2000web,lewandowski2015evaluating}. While using traditional cataloging, indexing and keyword matching techniques initially was sufficient for basic information retrieval on the Web, it was soon regarded to be a poor way to return search results when focusing on the commercialisation of Web content and search \cite{page1999pagerank}. With the entry of Google into the search engine market began the era of algorithms that take "advantage of the graph structure of the Web" to determine the popularity of Web content \cite{broder2000graph} in order to produce better, more relevant search results. Over time Google outperformed other search engine providers to become the market leading search engine, now with a market share of just over 74\% in 2017, getting as high as 90\% for mobile users thanks to the Chrome application that is embedded into the Android operating system for mobile devices. It is due to this widespread use of Google that the search engine is likely to play a role not only in people's private life but also impacts their behaviour when they are at work \cite{mangles2018statistics}.

\subsection{Search results personalisation}

Personalisation, regarded as a process that "tailors certain offerings (such as content, services, product recommendations, communications, and e-commerce interactions) by providers (such as e-commerce Web sites) to consumers (such as customers and visitors) based on knowledge about them, with certain goal(s) in mind" \cite{adomavicius2005personalization} was introduced to Web search by Google in 2005 as a means of getting better at providing the most relevant results \cite{google2005, google2009}. From the perspective of the search engine provider this was necessary since the vast (and continuously growing) amount of information available on the Web meant that more effective information retrieval systems were required in order to provide users the most relevant items according to their query \cite{brin2012reprint}. While Google's personalisation process is not fully transparent \cite{ormen2016googling}, it is known to include a plethora of behavioral signals captured from search engine users, such as past search results a user has clicked through, geographic location or visited Web sites, for example \cite{google2009,roesner2012detecting}. 

Such search result personalisation has led to concerns about what has been coined the \textsl{filter bubble}, i.e. the idea that people only read news that they are directly interested in and agree with, resulting in less familiarity with new or opposing ideas \cite{pariser2011filter,foster2012news}. However, there is still no academic consensus about whether the filter bubble actually does exist at all, or whether it is an overstated phenomenon \cite{foster2012news,dutton2017search,haim2018burst}. Hence, research as the one described here is still required to bring clarity to the current ambiguity about that matter.

\subsection{Related studies on search result comparison}
 
In \cite{du2011academic} a heavy reliance on search engines by academic users was found. This brought personalisation into the focus of research, prompting Salehi et al.'s \cite{salehi2015examining} research into personalisation of search results in academia. Using alternative search setups involving Startpage and Tor to depersonalize search results and comparing the rank order of different search results using the percentage of result overlap and Hamming distance \cite{salehi2015examining}, it was found that on average only 53\% of search results appear in both personalised and unpersonalised search.

The work by Hannak et al. \cite{hannak2013measuring} introduced a different approach for measuring personalisation of search results. They compared search results of a query performed by a participant (personalized) with the same query performed on a 'fresh' Google account (control) with no history. Comparison between the two sets is done using the Jaccard Index and Damerau-Levenshtein distance as well as Kendall's Tau to understand the difference in the rank order between two search results. They observed measurable personalisation when conducting search when signed into a Google account, and location personalisation from the use of the IP address. Ultimately, they observed that 11.7\% of search results were different due to personalisation.
 
In their audit of the personalisation and composition of politically related search result pages, Robertson et al. \cite{robertson2018auditing} found, while relatively low, a higher level of personalisation for participants who were signed into their Google account and/or regularly used Alphabet products. In order to account for the behavioral pattern of search engine users to stronger focus on top results \cite{lu2016effect} they used Rank-Biased Overlap \cite{webber2010similarity} to compare search results. 

Overall, these previous studies confirm corporate statements by Google regarding the use of location data and the profile of the user conducting the search \cite{google2005,google2009} for the tailoring of search results. Our work benefits from the continuous improvement of the methodologies used for search result comparison and transfers such a study setup into the public sector to shed light on the impact of personalisation of professional knowledge work.

\section{Research design and data}

We based the research design of our experiment on the previous studies that sought to investigate search result personalisation in an academic search context \cite{du2011academic,salehi2015examining} and the quantification of search result personalisation \cite{hannak2013measuring,robertson2018auditing}. Additionally, we introduce search result relevance as an additional dimension. The idea of self-assessed relevance has been explored perhaps most notably in \cite{pan2007google}. In this work we will investigate the relevance of the results that appear in personalised and unpersonalised search. 

\subsection{Study Participants}

We recruited 30 participants from the public sector following the typical procedure of convenience sampling (21 self-identified as female and 9 as male). Of these participants, 5 were at managerial level or higher. The results are slightly skewed towards one public sector organisation, with over half of the participants (16) from that particular organization, but participants were chosen randomly. Most participants are experienced, indicating they have been in their current industry for 10 years or more.

\subsection{Survey}

To gauge how 'important' the use of Google search was to public sector officials we performed a pre-experiment survey. The survey design was informed by two of the studies mentioned earlier of how academic researchers sought information by Du and Evans \cite{du2011academic}, and personalisation in academic search by Salehi et al. \cite{salehi2015examining}. Due to time constraints, the survey remained pre-qualifying, we did not perform a follow-up survey or interview. In order to determine the importance of Google and search engines, questions were directed at how important the participants believed that Google was to their work functions. For example, the survey included questions that seek to determine the extent of a participant's self-assessed reliance on Google and how often they used Google as part of their work routines.

\subsection{Experiment}
Following completion of the survey, we asked participants to perform two Google searches on their work computers, to simulate a "normal" search query that they might perform in the course of their everyday work duties. For each query that was performed, we performed the same search queries at the same time under two different conditions, both designed to obfuscate Google's knowledge of who performed the search. For the first query (Query 1), participants were asked to search something that they had actually searched before. For the second query (Query 2), participants were asked to search something that they would potentially search in the course of their work duties, but to the best of their knowledge had not searched before. This results in two queries being performed under 3 different search conditions.

\begin{figure}[htbp]
\centering
\includegraphics[width=0.4\textwidth]{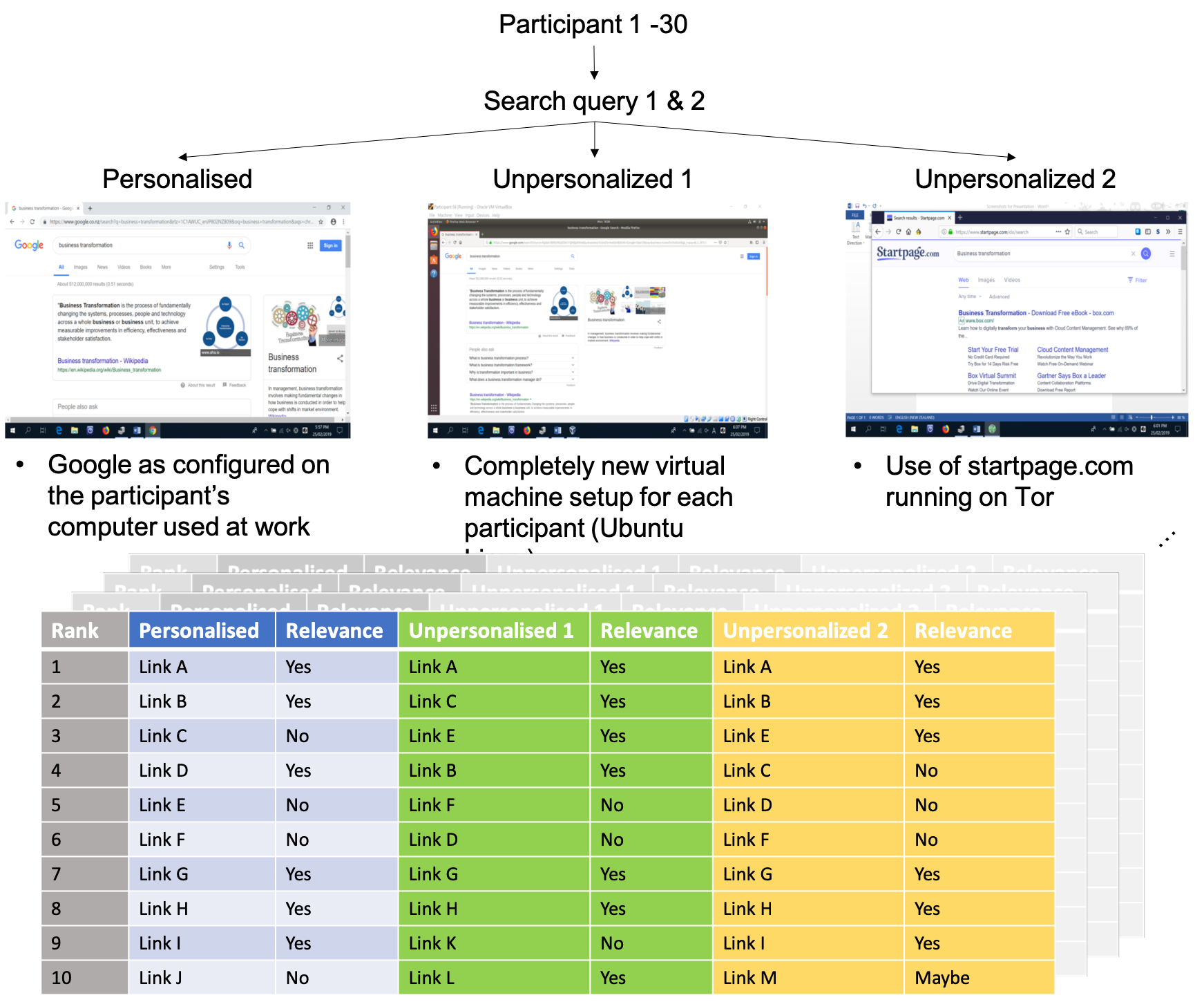}
\caption{Overview of the study setup and the three search result sets generated per query and participant.}
\normalsize
\label{fig:setup}
\end{figure}

\paragraph{Personalized search:} Participants performed the search for both queries at work on their work computers to simulate search performed during the course of their normal working day.

\paragraph{Unpersonalised search 1:} This condition attempted to depersonalise search query results through the use of a virtual machine running Mozilla Firefox on Linux’s Ubuntu Operating System. A virtual machine allows for a virtual computer to be created within a computer. By using a virtual machine, it is less likely that a person’s real identity will be left, unless they did something that allowed for their identity to be linked to the virtual machine \cite{van2017deviating}. The identity of whoever is performing the Google search should be tied to the virtual machine. Since each virtual machine was created for the purpose of this experiment, there is no history of any past searches that could influence the output of the search results, nor any identity to link to. A test run of this condition found that location personalisation was present, but only to the extent that the country from which the search was performed could be identified. It is believed that this is the extent of personalisation for this condition.

\paragraph{Unpersonalised search 2:} This second condition attempted to completely depersonalise search query results through the use of the Startpage search engine running on Tor, and is borrowed directly from Salehi et al. \cite{salehi2015examining}. Startpage is a search engine that gathers the best Google results, but does not reveal or store any personal information of the user. It has also been awarded a European Privacy Seal \cite{salehi2015examining}. Tor is essentially a modified Mozilla Firefox Browser with additional proxy applications and extensions that hides any identifying information by ‘fragmenting’ the links between the client and server by redirecting the traffic through thousands of relays \cite{van2017deviating}.

After each search result was retrieved, we asked the participants to rate the relevance of each of the top 10 search result items on a three-point Likert scale (relevant, maybe relevant, not relevant). This three-point scale was chosen to reduce potential ambiguity of more nuanced levels on any larger scale with respect to the rating of the relevancy of search results. When using larger scales we experienced higher variance in how study participants interpret the different levels which would lead to undesired limitations for the study of the result relevancy.

\subsection{Data analysis}

The survey responses as well as the self-assessed relevance scores for search results were analysed using exploratory data analysis (EDA) techniques such as calculating the mean and standard deviation (SD) for survey responses. To compare the rank of any pair of ordered sets of URIs $A$ and $B$ we use the Rank-Biased Overlap (RBO) measure as justified in \cite{robertson2018auditing}. RBO provides a rank similarity measure that takes top-weightedness (stronger penalties for differences at the top of a list) and incompleteness (lists with different items) into account. In our study setup we only compare sets of equal size limited to the top 10 results retrieved under the three aforementioned query conditions.


The RBO measure contains the parameter $\Psi \in [0,1]$, which represents the degree to which a fictive user is focused on top ranked results, with smaller values of $\Psi$ reflecting stronger focus on top results. RBO is a common measure for this kind of analysis and outperforms other measures to assess the similarity or distance of vectors of strings (e.g. hamming distance) due to the possibility to factor in the focus on top results. 

For each unpersonalised result set (i.e. unpersonalised search 1 and unpersonalised search 2), we computed the proportion of URIs self-assessed as relevant that were not in the respective personalised search result set. This provides us with an understanding how much relevant information is missed in the personalised search. 

We also computed six sets of URIs that were common between pairs of result sets (leading to three such sets per query) but that were rated differently in terms of their relevance in order to find out whether participants were consistent with their relevance assessment. To investigate deeper whether the rank order may add bias to the participants' self-assessment of the search result relevance, we also analysed the rank change for URIs within those sets (i.e. whether a URI that was assessed differently moved up or down in the ranking).

Finally, we derived the sets of URIs that are deemeed relevant in any of the unpersonalised result sets but that were not present in a respective personalised search result. To understand whether there is any bias in the participant's assessment of the relevancy (e.g. implicit assumption that highly ranked results in search must be relevant) we then computed the distribution of the ranks of those URIs in the four respective unpersonalised search result sets.

\section{Findings}

\subsection{Trust in and reliance on Google in the public sector}

As presented in Table~\ref{tab:use} the majority of participants indicated that they use Google every day for both work and non-work purposes. Furthermore, most participants said that Google is their first point of enquiry as opposed to other sources such as asking co-workers. Participants also indicated that they do not compare the results of their Google searches with other search engines. These responses indicate a high level of trust in and reliance on Google in the public sector. The responses to questions asking about the quality of people's work if they were not able to use Google further confirms this reliance. Participants indicated that they generally believed that their work would become of worse quality if they could not use Google, even if they could use other sources of information.

\begin{table}[hbt]
\centering
\scriptsize
\begin{tabular}{l|l}
\textbf{Survey item}                        & \textbf{Mean response (SD)} \\ \hline
Frequency of use                   & 4 (0.92)                                        \\ \hline
As the first point of enquiry      & 4 (0.91)                                        \\ \hline
Search engine comparison frequency & 2 (1.05)                                        \\ \hline
Impact on quality of work          & 4 (0.90)                                        \\
\end{tabular}
\caption{Mean and standard deviation for the survey responses related to the use and trust in Google as a first and single point of online research.}\label{tab:use}
\end{table}

That the overwhelming majority of participants use Google as their first point of enquiry at work draws comparisons with studies that found that around 80\% of Internet users in an academic context used Google search as their first point of enquiry \cite{du2011academic,salehi2015examining}. The participants in the study by Du et al. \cite{du2011academic} indicated that this was because they found Google easy to use, and that it had become a habit to use Google as the first option when they needed to search for information. While participants in our study were not explicitly asked why they used Google as their first point of enquiry, other factors such as the fact that they did indicate that they do not compare results with other search engines' results point into the direction that Google plays a similar role in the public sector.

\subsection{Variance in personalised and unpersonalised search results}

\begin{figure}[h]
     \centering
     \begin{subfigure}[b]{0.25\textwidth}
         \centering
         \includegraphics[width=\textwidth]{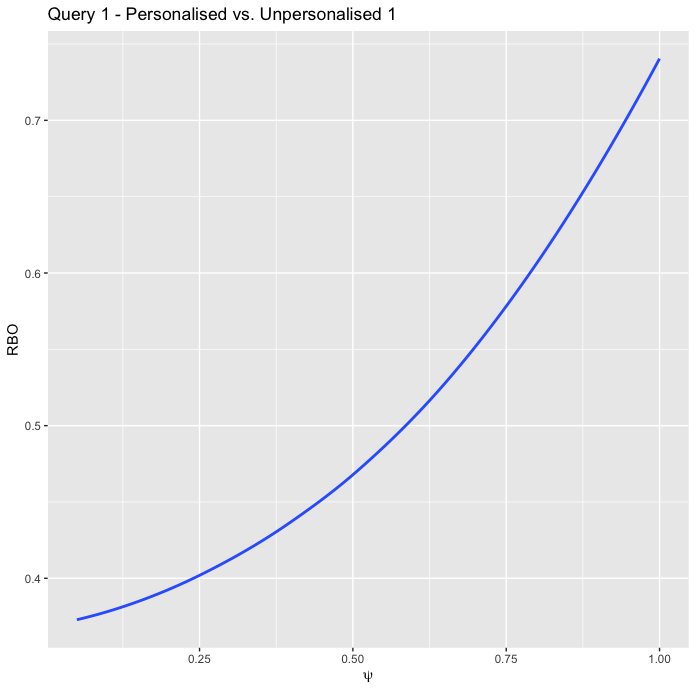}
     \end{subfigure}
     \hfill
     \begin{subfigure}[b]{0.25\textwidth}
         \centering
         \includegraphics[width=\textwidth]{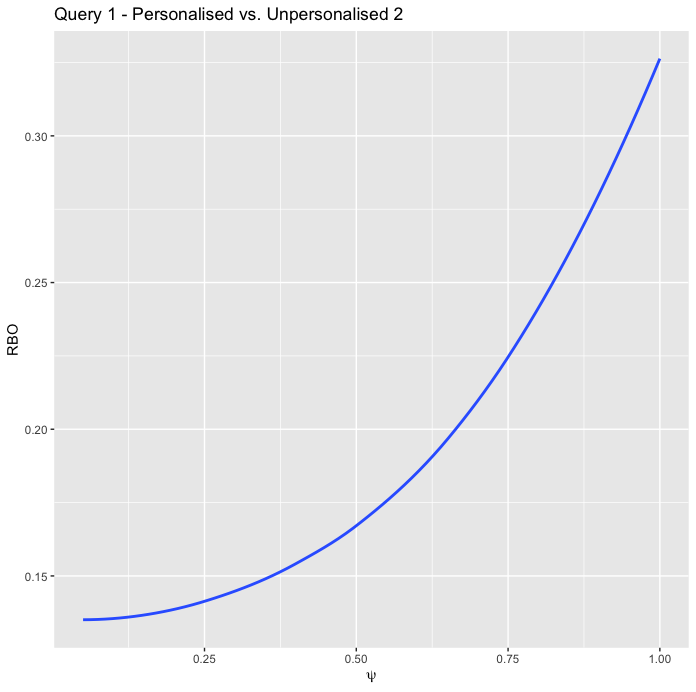}
     \end{subfigure}
     \hfill
     \begin{subfigure}[b]{0.25\textwidth}
         \centering
         \includegraphics[width=\textwidth]{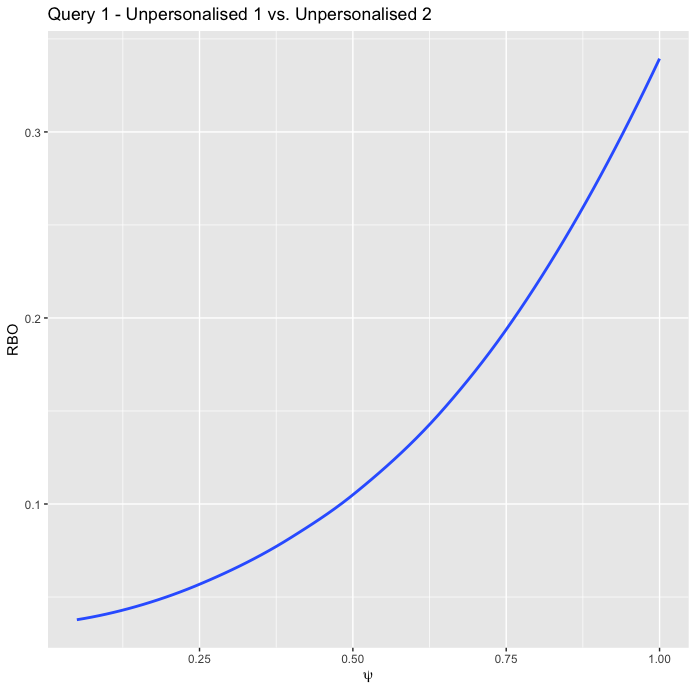}
     \end{subfigure}
     
        \begin{subfigure}[b]{0.25\textwidth}
         \centering
         \includegraphics[width=\textwidth]{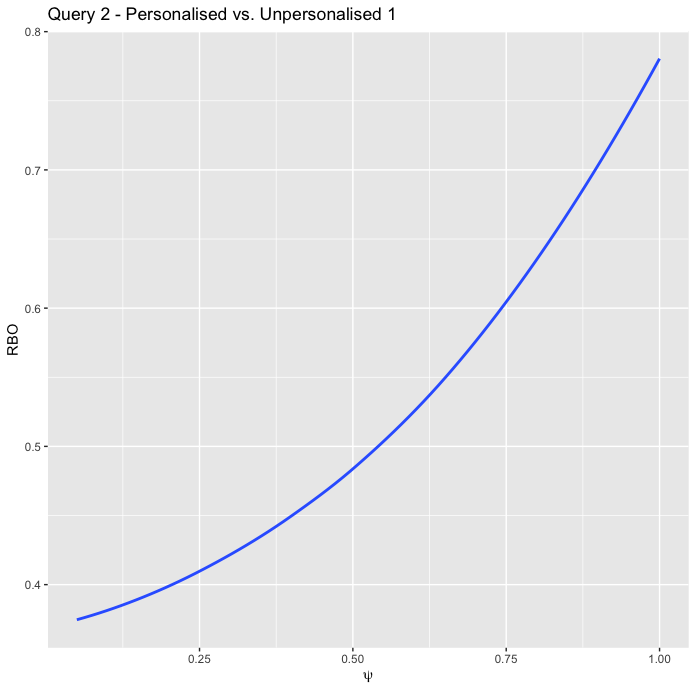}
     \end{subfigure}
     \hfill
     \begin{subfigure}[b]{0.25\textwidth}
         \centering
         \includegraphics[width=\textwidth]{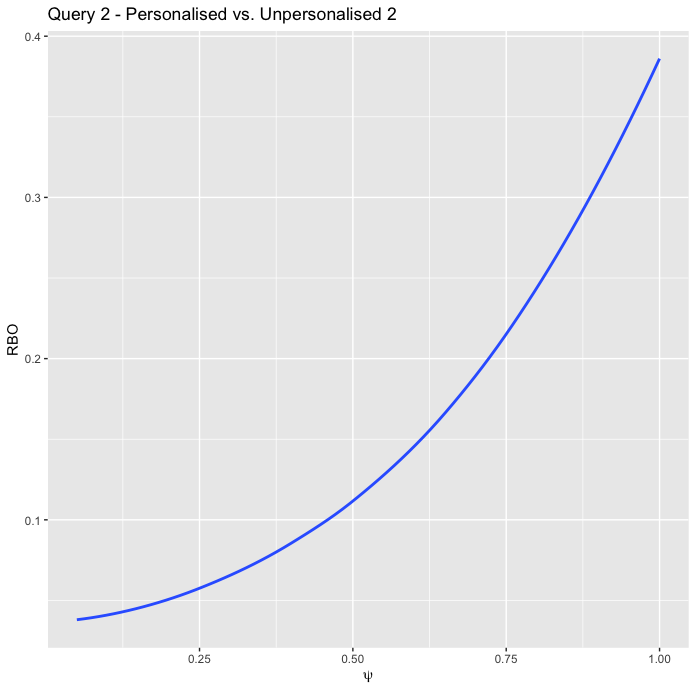}
     \end{subfigure}
     \hfill
     \begin{subfigure}[b]{0.25\textwidth}
         \centering
         \includegraphics[width=\textwidth]{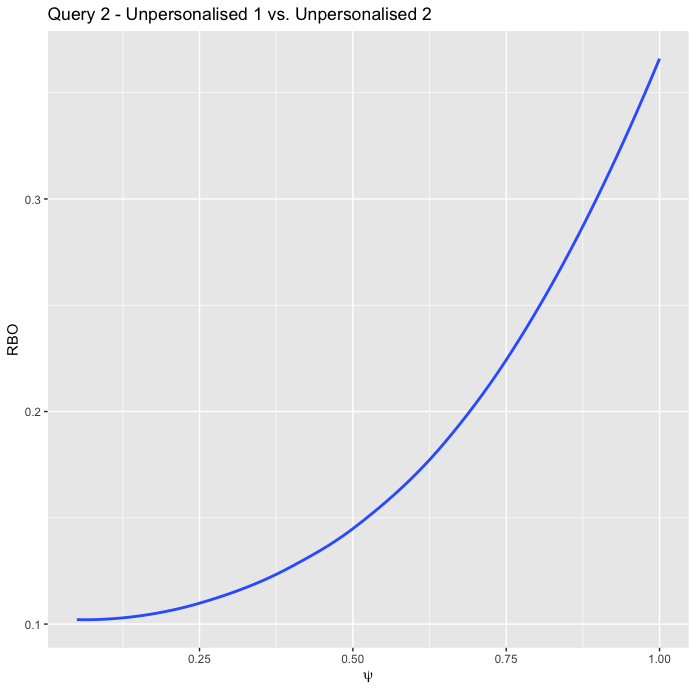}
     \end{subfigure}
        \caption{Rank-Biased Overlap analysis with variable $\Psi$ threshold from 0 to 1 (in steps of 0.05) for both queries performed by the study participants under all three experimental conditions.}
        \label{fig:rbo}
\end{figure}

Figure~\ref{fig:rbo} shows the results of our analysis of the RBO. We plotted a smoothed line graph for $20$ RBO scores for $\Psi$ in the range from $0.05$ to $1.0$ (increased in steps of $0.05$). Since smaller $\Psi$ values indicate stronger focus on top ranks in search results, the shape of these graphs shows that the similarity of search results is consistently lower for top ranked search results and increases as lower ranks are taken into account. The similarity of search results is consistently the highest for personalised and unpersonalised search 1, reaching an RBO of almost $0.8$ when focusing on low-ranked results and a lower bound of around $0.4$ when relaxing the $\Psi$ parameter to focus on the top results only. Any comparison with unpersonalised search 2 does not even reach an RBO of $0.4$ even when focusing on low-ranked results.

While this result supports the recommendations to use advanced measures to compare search result rankings \cite{webber2010similarity,robertson2018auditing}, we suggest that it is a call for deeper investigations into the lower ranks of search results to quantify and qualify the information professional knowledge workers would miss out on if they focus on top ranked search results.

\subsection{Result relevance in personalised and unpersonalised search}

With respect to the search result relevance assessment we find that between half and two-thirds of the search results have been assessed as relevant by the study participants. However, the mean number of maybe responses for query 2 result sets is about 50\% smaller than that of query 1, which means participants were making more certain assessments whether a result is relevant or not for the second query they performed during the experiment.

\begin{table}
\centering
\scriptsize
\arrayrulecolor{black}
\begin{tabular}{!{\color{black}\vrule}l!{\color{black}\vrule}l!{\color{black}\vrule}l!{\color{black}\vrule}l!{\color{black}\vrule}l!{\color{black}\vrule}} 
\hline
\multicolumn{2}{!{\color{black}\vrule}l!{\color{black}\vrule}}{} & Yes (SD) & No (SD) & Maybe (SD)  \\ 
\hline
\multirow{3}{*}{Query 1} & Personalised                          & 6 (2.6)                    & 2.6 (2.5)                 & 1.3 (1.7)                     \\ 
\cline{2-5}
                         & Unpersonalised 1                      & 6 (2.7)                    & 2.9 (2.7)                 & 1.1 (1.7)                     \\ 
\cline{2-5}
                         & Unpersonalised 2                      & 5 (2.8)                    & 3.8 (2.9)                 & 1.2 (1.6)                     \\ 
\hline
\multirow{3}{*}{Query 2} & Personalised                          & 6.6 (2.1)                  & 2.8 (2.2)                 & 0.6 (1.3)                     \\ 
\cline{2-5}
                         & Unpersonalised 1                      & 6.3 (2.5)                  & 3.1 (2.5)                 & 0.5 (1.1)                     \\ 
\cline{2-5}
                         & Unpersonalised 2                      & 6.2 (2.5)                  & 3.3 (2.6)                 & 0.5 (1.4)                     \\
\hline
\end{tabular}
\arrayrulecolor{black}
\caption{Means for the yes, no and maybe responses of the relevance assessment.}\label{tab:surv}
\end{table}


The proportion of URIs that are found in different result sets for the same participant but that this participant rated differently ranges from 15.7\% up to 19.7\% of all URIs in the intersection of pairs of result sets as shown in Table~\ref{tab:uris}. We also highlight that this relevance assessment inconsistency is higher for query 1, which is the query that the participant did perform before as part of her work.

\begin{table}
\centering
\scriptsize
\arrayrulecolor{black}
\begin{tabular}{!{\color{black}\vrule}l!{\color{black}\vrule}l!{\color{black}\vrule}l!{\color{black}\vrule}} 
\hline
\multirow{3}{*}{Query 1} & Personalised vs. unpersonalised 1     & 19.7\%                                                                                                              \\ 
\cline{2-3}
                         & Personalised vs. unpersonalised 2     & 19.7\%                                                                                                              \\ 
\cline{2-3}
                         & Unpersonalised 1 vs. unpersonalised 2 & 19.3\%                                                                                                              \\ 
\hline
\multirow{3}{*}{Query 2} & Personalised vs. unpersonalised 1     & 18\%                                                                                                                \\ 
\cline{2-3}
                         & Personalised vs. unpersonalised 2     & 16.3\%                                                                                                              \\ 
\cline{2-3}
                         & Unpersonalised 1 vs. unpersonalised 2 & 15.7\%                                                                                                              \\
\hline
\end{tabular}
\arrayrulecolor{black}
\caption{Proportion of inconsistently rated URIs.}\label{tab:uris}
\end{table}


\begin{figure}
     \centering
     \begin{subfigure}[b]{0.25\textwidth}
         \centering
         \includegraphics[width=\textwidth]{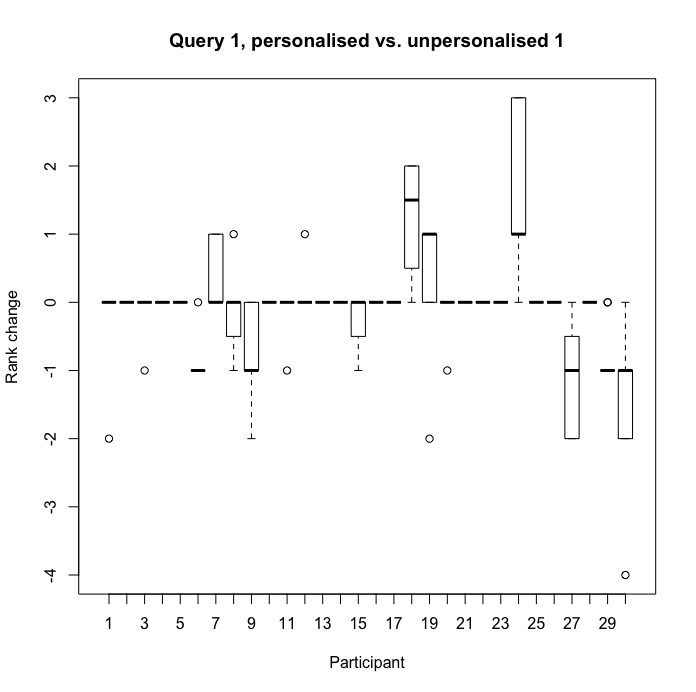}
     \end{subfigure}
     \hfill
     \begin{subfigure}[b]{0.25\textwidth}
         \centering
         \includegraphics[width=\textwidth]{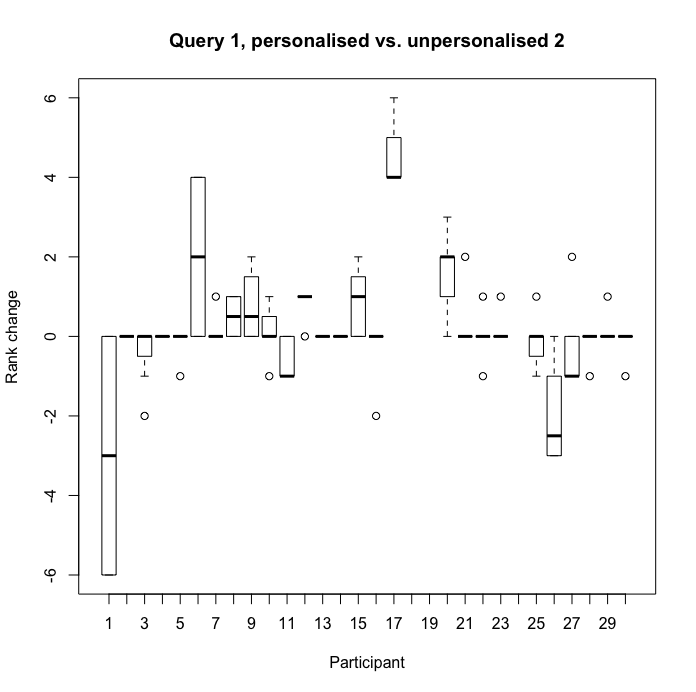}
     \end{subfigure}
     \hfill
     \begin{subfigure}[b]{0.25\textwidth}
         \centering
         \includegraphics[width=\textwidth]{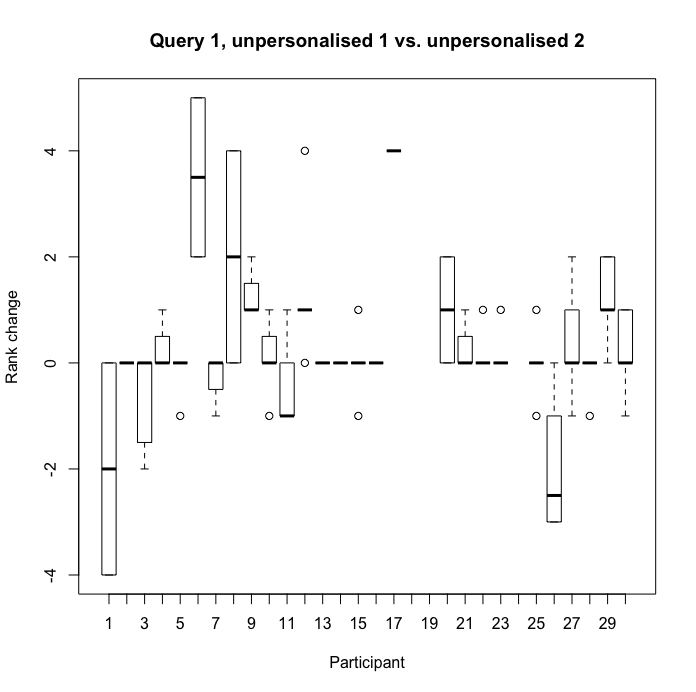}
     \end{subfigure}
        
        \begin{subfigure}[b]{0.25\textwidth}
         \centering
         \includegraphics[width=\textwidth]{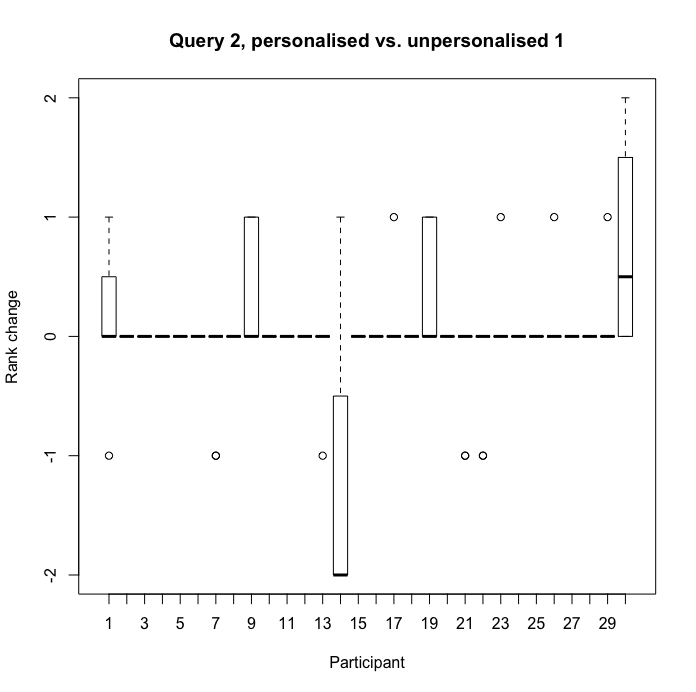}
     \end{subfigure}
     \hfill
     \begin{subfigure}[b]{0.25\textwidth}
         \centering
         \includegraphics[width=\textwidth]{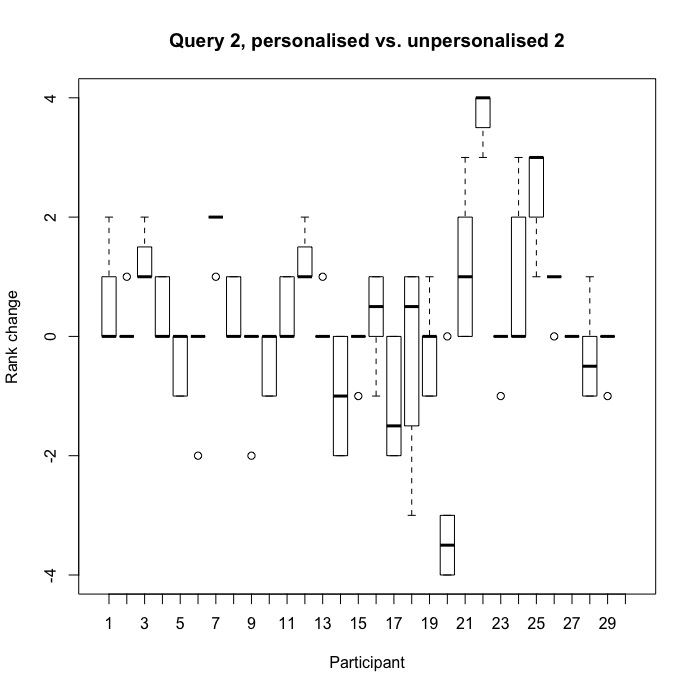}
     \end{subfigure}
     \hfill
     \begin{subfigure}[b]{0.25\textwidth}
         \centering
         \includegraphics[width=\textwidth]{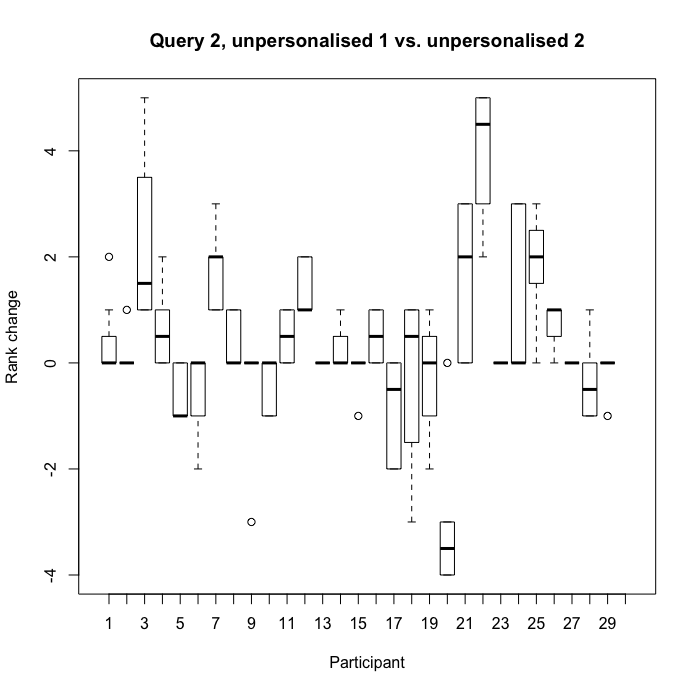}
     \end{subfigure}
        \caption{Change in rank of unique URIs for which the users' relevance assessment varied between the three experimental conditions.}
        \label{fig:boxplot1}
\end{figure}

The results of our deeper investigation of whether unique URIs for which the participants' assessment varied between the different conditions are depicted in Figure~\ref{fig:boxplot1}. The graphs show that macroscopically there is no tendency towards URIs that are ranked higher or lower to be assessed inconsistently. 

\begin{figure}[htbp]
\centering
\includegraphics[width=0.4\textwidth]{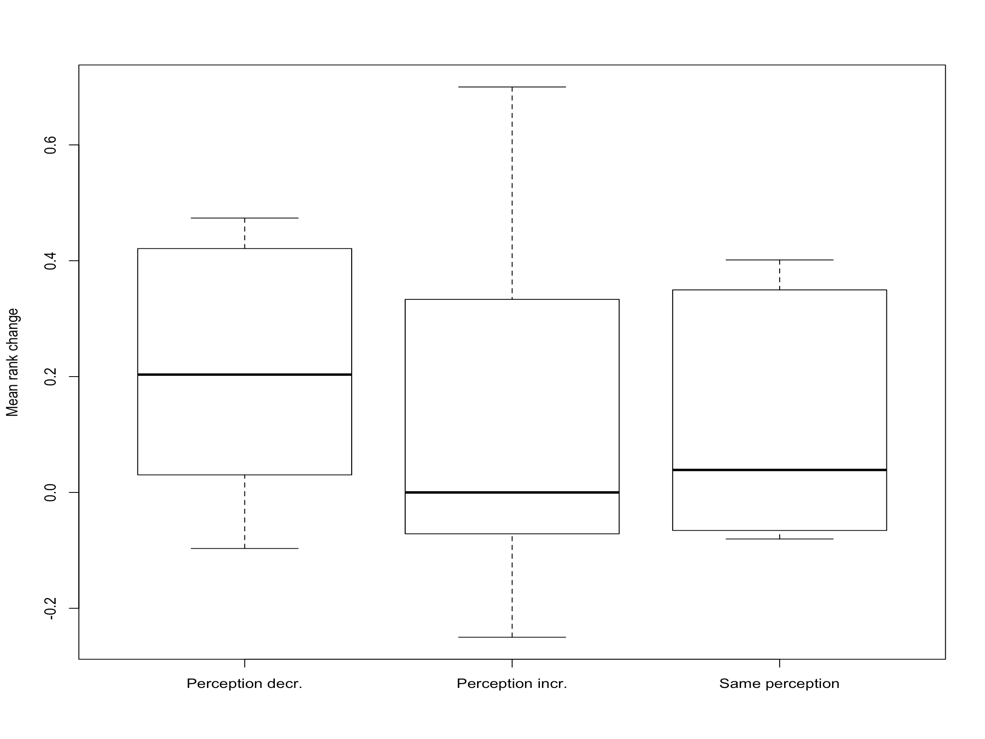}
\caption{Change in rank of unique URIs for which the users' relevance assessment increased, decreased or remained the same.}
\normalsize
\label{fig:rankchange}
\end{figure}

Further to the results shown in Figure~\ref{fig:boxplot1} we also investigated whether there is any trend towards higher or lower ranking of inconsistently assessed URIs specifically for those URIs for which the assessment increased (e.g. assessment of maybe to yes), decreased (e.g. assessment from yes to no) or remained the same. As Figure~\ref{fig:rankchange} shows, there is if at all a moderate tendency that URIs are ranked higher if the perception changed bu that it has no influence whether the perception increased or decreased.

All these results related to the relevance assessment can be interpreted that the self-assessment of search result relevance is either a task prone to human error or that the participants are impacted by an unobserved factor in the experimental setup that causes this behavior. The former would be again in-line with previous studies with regards to Internet search behaviour and people's ability to assess search result relevance \cite{pan2007google}, while the latter means we additionally suggest that there is potentially a cognitive bias at work impacting the participants' assessment. In other words, the observation that the relevance assessment inconsistency is higher for query 1, which is the query that the participant did perform before as part of her work, allows to question whether the links in the query 1 result sets were more tricky to assess consistently because of the fact that this was a query they performed before so had more detailed knowledge about the topic leading to more nuanced opinions, or whether the participants just became more certain in how to rate relevance as the experiment progressed because of a training effect kicking in.

\subsubsection{Missing relevant results}

Table~\ref{tab:uris2} shows the proportion of unique URLs that were exclusively found in unpersonalised search result sets but considered relevant as per participants' assessment. The numbers show that in both, unpersonalised search 1 and unpersonalised search 2, there is a significant amount of relevant information to be found. Most significantly, the  depersonalised search setup using Tor and startpage.com allowed to retrieve up to 20.3\% of relevant information that were not found under personalised search conditions in our experiment. While previous studies also found that people may miss information due to search engine personalisation \cite{hannak2013measuring}, our unique experiment using the two different unpersonalised search settings allows to further detail how one may circumvent this filter bubble effect and also quantifies the difference this may make.

\begin{table}
\centering
\scriptsize
\arrayrulecolor{black}
\begin{tabular}{!{\color{black}\vrule}l!{\color{black}\vrule}l!{\color{black}\vrule}l!{\color{black}\vrule}} 
\hline
\multirow{2}{*}{Query 1} & Unpersonalised 1 & 7.3\%                                                                                                                                                                     \\ 
\cline{2-3}
                         & Unpersonalised 2 & 16.7\%                                                                                                                                                                    \\ 
\hline
\multirow{2}{*}{Query 2} & Unpersonalised 1 & 6\%                                                                                                                                                                       \\ 
\cline{2-3}
                         & Unpersonalised 2 & 20.3\%                                                                                                                                                                    \\
\hline
\end{tabular}
\arrayrulecolor{black}
\caption{Overall proportion of unique URLs that are not found in personalised search but in one of the unpersonalised searches and that are assessed as relevant.}\label{tab:uris2}
\end{table}

\begin{figure}[h]
     \centering
     \begin{subfigure}[b]{0.22\textwidth}
         \centering
         \includegraphics[width=\textwidth]{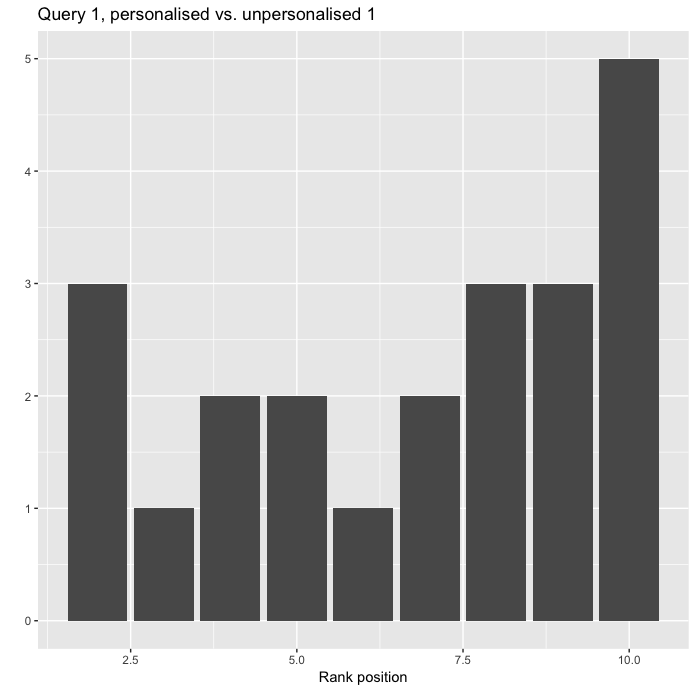}
     \end{subfigure}
     \hfill
     \begin{subfigure}[b]{0.22\textwidth}
         \centering
         \includegraphics[width=\textwidth]{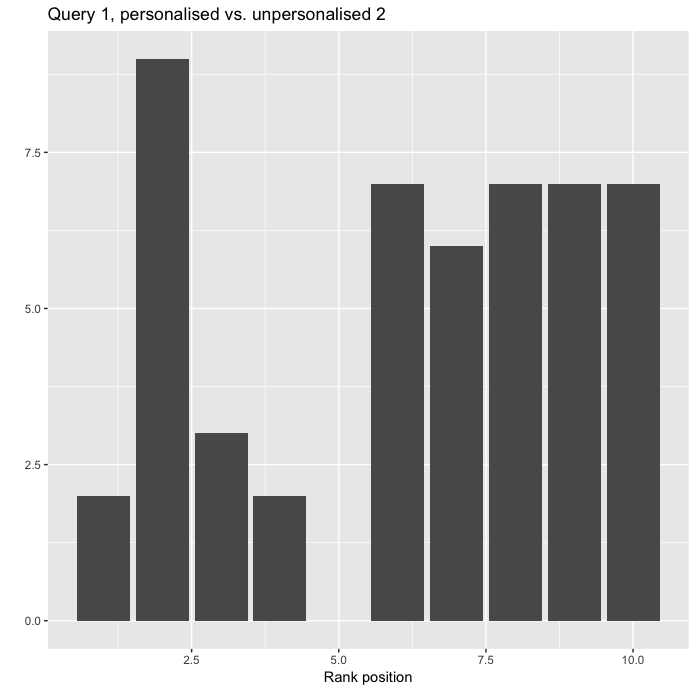}
     \end{subfigure}
     \hfill
     \begin{subfigure}[b]{0.22\textwidth}
         \centering
         \includegraphics[width=\textwidth]{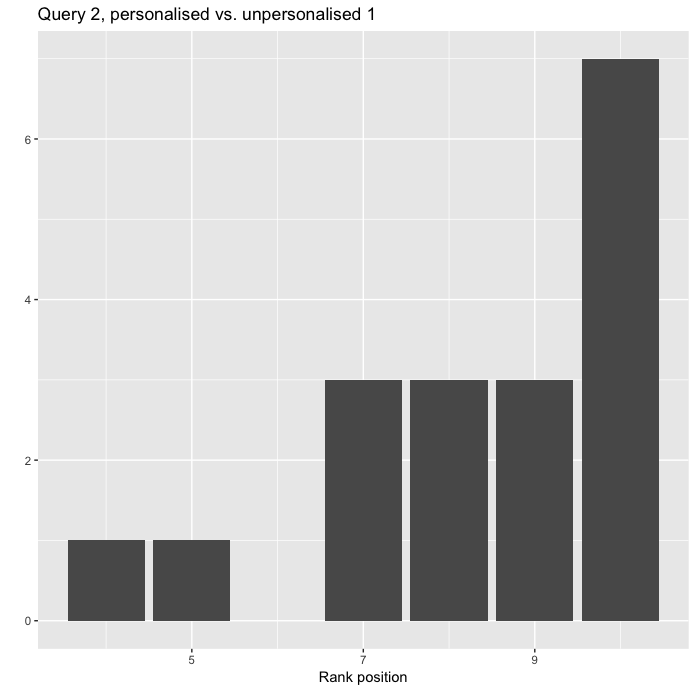}
     \end{subfigure}
     \hfill
     \begin{subfigure}[b]{0.22\textwidth}
         \centering
         \includegraphics[width=\textwidth]{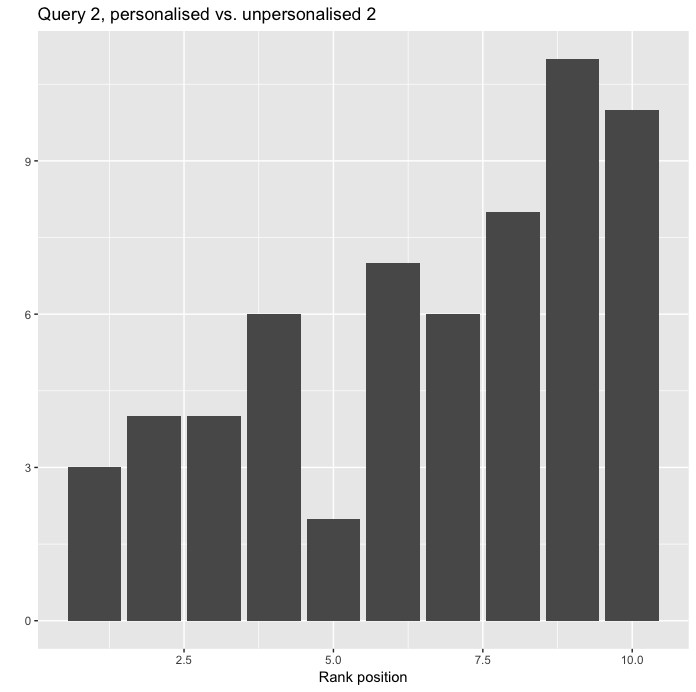}
     \end{subfigure}
        \caption{Rank order distribution of links that are missing from personalised but are deemed relevant in any of the unpersonalised search result sets.}
        \label{fig:relevantorder}
\end{figure}

Figure~\ref{fig:relevantorder} shows the rank order distribution of those relevant results that are missing from personalised search. The distributions show a weak tendency that those missing but relevant information are found in the lower ranks of search results. In the light of multiple previous studies that found that search engine users focus substantially on top ranked results \cite{granka2004eye,pan2007google} this is important because it means finding all relevant information is not just a challenge to be solved by either removing or circumventing personalisation algorithms but also a user interface (UI) and user experience (UX) design issue.

\section{Limitations}

Similar to previous work \cite{hannak2013measuring,salehi2015examining} our research was limited due to the small sample size. This was a practical constraint due to the way the experiment was designed and can only be avoided when either accepting uncertainty about whether the participants are actually public sector workers (e.g. by running it as a self-administered online experiment) or by running it over a much longer period of time. We consider the latter for our future research combined with an extension to cover alternative search engines and performing the experiment in multiple countries to also account for localisation. Future research should also expand the investigation of relevance of search results and in particular the properties and implications of self-assessment of relevance. Relevance is a subjective matter, and how the participants rated relevance in our experiments differed between each participant. Pan et al. \cite{pan2007google} were able to take this subjectivity into account through an objective third party evaluation, which we did not do, because our participants were the subject matter experts for the queries that they performed in a work context.

\section{Conclusion}

In this paper we presented findings from a Web search experiment involving public sector workers. We investigated not only how important they perceive Google Web search for fulfilling their information needs, but also whether Google's Web search personalisation means they may miss relevant information. We find that the majority of participants in our experimental study are neither aware that there is a potential problem nor do they have a strategy to mitigate the risk of missing relevant information when performing online searches. Most significantly, we provide empirical evidence that up to $20\%$ of relevant information may be missed due to Web search personalisation.

The fact that personalisation has an impact on search results was not surprising, particularly in the light of previous studies focused on academic Web search \cite{hannak2013measuring,salehi2015examining}. However, our work provides new empirical evidence for this phenomenon in the public sector. Therefore, our research has significant implications for public sector professionals, who should be provided with training about the potential algorithmic and human biases that may affect their judgments and decision making, as well as clear guidelines how to minimise the risk of missing relevant information. This does not just involve comparing search results using different search engines and to actively look further down the ranks for relevant search results, but maybe even that it is necessary that public sector agencies provide dedicated infrastructure to obfuscate the users' identities to circumvent personalisation.


%
%
%
 \bibliographystyle{splncs04}
 \bibliography{literature}

\end{document}